\documentclass{article}
\usepackage{amssymb}
\usepackage{latexsym}
\usepackage{amsmath}
\usepackage{graphicx}
\usepackage{float}

\newlength{\dinwidth}
\newlength{\dinmargin}
\begin{document}



\thispagestyle{empty} \vspace*{1cm} 
\vspace*{2cm}

\begin{center}
{\LARGE Dissipative quantum mechanics and Kondo-like impurities on
noncommutative two-tori }

{\LARGE \ }

{\large Patrizia Iacomino\footnote{{\large {\footnotesize Dipartimento di
Matematica e Applicazioni ''R. Caccioppoli'',}\textit{\ {\footnotesize %
Universit\'{a} di Napoli ``Federico II''}, }{\small Compl.\ universitario M.
Sant'Angelo, Via Cinthia, 80126 Napoli, Italy}}}, Vincenzo Marotta\footnote{%
{\large {\footnotesize Dipartimento di Scienze Fisiche,}\textit{\
{\footnotesize Universit\'{a} di Napoli ``Federico II''\ \newline
and INFN, Sezione di Napoli}, }{\small Compl.\ universitario M. Sant'Angelo,
Via Cinthia, 80126 Napoli, Italy}}},} {\large Adele Naddeo\footnote{{\large
{\footnotesize Dipartimento di Fisica \textit{''}E. R. Caianiello'',}\textit{%
\ {\footnotesize Universit\'{a} degli Studi di Salerno and CNISM, Unit\'{a}
di Ricerca di Salerno,}}{\small Via Ponte Don Melillo, 84084 Fisciano (SA),
Italy}}}}

{\small \ }

\textbf{Abstract\\[0pt]
}
\end{center}

\begin{quotation}
In a recent paper, by exploiting the notion of Morita equivalence for field
theories on noncommutative tori and choosing rational values of the
noncommutativity parameter $\theta $ (in appropriate units), a general
one-to-one correspondence between the $m$-reduced conformal field theory
(CFT) describing a quantum Hall fluid (QHF) at paired states fillings $\nu =%
\frac{m}{pm+2}$ \cite{cgm2,cgm4} and an Abelian noncommutative field theory
(NCFT) has been established \cite{AV2}. That allowed us to add new evidence
to the relationship between noncommutativity and quantum Hall fluids\cite%
{ncmanybody}. On the other hand, the $m$-reduced CFT is equivalent to a
system of two massless scalar bosons with a magnetic boundary interaction as
introduced in \cite{maldacena}, at the so called ``magic''\ points. We are
then able to describe, within such a framework, the dissipative quantum
mechanics of a particle confined to a plane and subject to an external
magnetic field normal to it. Here we develop such a point of view by
focusing on the case $m=2$ which corresponds to a quantum Hall bilayer. The
key role of a localized impurity which couples the two layers is emphasized
and the effect of noncommutativity in terms of generalized magnetic
translations (GMT) is fully exploited. As a result, general GMT operators
are introduced, in the form of a tensor product, which act on the QHF and
defect space respectively, and a comprehensive study of their rich structure
is performed.

\vspace*{0.5cm}


{\footnotesize PACS: 11.25.Hf, 11.10.Nx, 73.43.Cd\newpage }\baselineskip%
=18pt \setcounter{page}{2}
\end{quotation}

\section{Introduction}

In the last years the physics of open strings and $D$-branes in constant
background electric and magnetic fields \cite{openstring1} has spurred a
renewed interest because it gives an explicit realization of old conjectures
about the nature of spacetime at very short distance scales. Indeed by using
string states as probes of the short distance structure it is clear that one
cannot probe lengths smaller than the intrinsic length of the strings. As a
consequence the structure of the spacetime must change below the string
scale in such a way that the spacetime coordinates become noncommuting
operators. Therefore noncommutative geometry has been proposed to describe
the short scale structure of spacetime and nonperturbative features of
string theory \cite{witten1}. On the other hand, certain configurations of $%
D $-branes in massive $IIA$ string theory have been proposed to describe
quantum Hall fluids as a realization of noncommutative Chern-Simons theories
\cite{brodie1}. That stimulated research on new effective theories in the
realm of high energy physics which could better describe condensed matter
phenomena and could help to clarify the general relationship between
noncommutativity and condensed matter physics \cite{ncmanybody}\cite{poly1}.

In order to better understand the role played by noncommutative field
theories (NCFT) \cite{ncft1} within the string theory context and its
relationship with the physics of QHF let us briefly recall the simplest
framework in which they arise in a natural way as an effective description
of the dynamics, namely the quantum mechanics of the motion of $N_{e}$
charged particles in two dimensions subjected to a transverse magnetic
field, known also as Landau problem \cite{landau}. The strong field limit $%
B\rightarrow \infty $ at fixed mass $m$ projects the system onto the lowest
Landau level and, for each particle $I=1,...,N_{e}$, the corresponding
coordinates ($\frac{eB}{c}x_{I},y_{I}$) are a pair of canonical variables
which satisfy the commutation relations $\left[ x_{I},y_{I}\right] =i\delta
_{I,J}\theta $, $\theta =\frac{\hbar c}{eB}\equiv l_{M}^{2}$ being the
noncommutativity parameter. In this picture the electron is not a point-like
particle and can be localized at best at the scale of the magnetic length $%
l_{M}$. The same thing happens to the endpoints of an open string attached
to a $D$-brane embedded in a constant magnetic field \cite{openstring1},
which is the string theory analogue of the Landau problem: the $D$-brane
worldvolume becomes a noncommutative manifold. In order to clarify such a
point of view let us consider the worldsheet field theory for open strings
attached to $D$-branes, which is defined by a $\sigma $-model on the string
worldsheet $\Sigma $ with action
\begin{equation}
S_{\Sigma }=\frac{1}{4\pi l_{s}^{2}}\int_{\Sigma }d^{2}\xi \left(
g_{ij}\partial ^{a}y^{i}\partial _{a}y^{j}-2\pi \mathrm{i}%
l_{s}^{2}B_{ij}\epsilon ^{ab}\partial _{a}y^{i}\partial _{b}y^{j}\right) .
\label{mb1}
\end{equation}
Here $y^{i}$ are the open string endpoint coordinates, $\xi ^{a}$, $a=1,2$
are local coordinates on the surface $\Sigma $, $l_{s}$ is the intrinsic
string length, $g_{ij}$ is the spacetime metric and $B_{ij}$ is the
Neveu-Schwarz two-form which is assumed non-degenerate and can be viewed as
a magnetic field on the $D$-brane. Indeed, when $B_{ij}$ are constant the
second term in Eq. (\ref{mb1}) can be integrated by parts and gives rise to
the boundary action:
\begin{equation}
S_{\partial \Sigma }=-\frac{\mathrm{i}}{2}\int_{\partial \Sigma
}dtB_{ij}y^{i}\left( t\right) \overset{.}{y}^{j}\left( t\right) ,
\label{mb2}
\end{equation}
where $t$ is the coordinate of the boundary $\partial \Sigma $ of the string
world sheet lying on the $D$-brane worldvolume and $\overset{.}{y}%
^{i}=\partial y^{i}/\partial t$. Such a boundary action formally coincides
with the one of the Landau problem in a strong field. Now, by taking the
Seiberg-Witten limit \cite{witten1}, i. e. by taking $g_{ij}\sim
l_{s}^{4}\sim \varepsilon \rightarrow 0$ while keeping fixed the field $%
B_{ij}$, the effective worldsheet field theory reduces to the boundary
action (\ref{mb2}) and the canonical quantization procedure gives the
commutation relations $\left[ y^{i},y^{j}\right] =i\theta ^{ij}$, $\theta =%
\frac{1}{B}$ on $\partial \Sigma $. Summarizing, the quantization of the
open string endpoint coordinates $y^{i}\left( t\right) $ induces a
noncommutative geometry on the $D$-brane worldvolume and the effective
low-energy field theory is a noncommutative field theory for the massless
open string modes.

In such a picture we can also consider the tachyon condensation phenomenon
which introduces the following boundary interaction for the open string:
\begin{equation}
S_{T}=-\mathrm{i}\int_{\partial \Sigma }dtT\left( y^{i}\left( t,0\right)
\right) ,  \label{mb3}
\end{equation}
where $T\left( y^{i}\left( t,0\right) \right) $ is a general tachyon
profile. In general, $D$-branes in string theory correspond to conformal
boundary states, i. e. to conformally invariant boundary conditions of the
associated CFT. They are charged and massive objects that interact with
other objects in the bulk, for instance through exchange of higher closed
string modes. In turn boundary excitations are described by fields that can
be inserted only at points along the boundary, i. e. the boundary fields.
There exists an infinite number of open string modes, which correspond to
boundary fields in the associated boundary CFT. In other words tachyonic
condensation is a boundary phenomenon and as such it is related to
Kondo-like effects in condensed matter systems \cite{kondo1}\cite{affleck1}.
In this case, the presence of the background $B$-field makes the open string
states to disappear and that explains the independence on the background
\cite{seiberg1}.

The issue of background independence also appears within the context of
gauge theories on noncommutative space and, in particular, on a
noncommutative two-torus where it is intimately related to Morita duality
\cite{morita}\cite{scw1}\cite{scw2}. Such a kind of duality establishes a
relation, via a one-to-one correspondence, between representations of two
noncommutative algebras and, within the context of gauge theories on
noncommutative tori, it can be viewed as a low energy analogue of $T$%
-duality of the underlying string model \cite{string1}. As such, it results
a powerful tool in order to establish a correspondence between NCFT and well
known standard field theories. Indeed, for rational values of the
noncommutativity parameter, $\theta =\frac{1}{N}$, one of the theories
obtained by using the Morita equivalence is a commutative field theory of
matrix valued fields with twisted boundary conditions and magnetic flux $c$
\cite{wilson1} (which, in a string description, behaves as a $B$-field
modulus).

On the other hand, in Ref. \cite{cgm5} the effect of the dissipative term ($%
\eta \overset{{\cdot }}{q}$) on the motion of an electron confined in a
plane and in the presence of an external magnetic field $B$, normal to the
plane, was analyzed. By using the correspondence principle it was possible
to quantize the system and to study its time evolution on large time scales (%
$t\gg \frac{1}{\eta }$) employing coherent states techniques. It was found
that the effect of dissipation would simply be accounted for by a rotation
combined with a scale transformation on the coordinate $z$ of the electron: $%
z\rightarrow (\rho e^{i\gamma })z$, $\rho =\left( \frac{\eta ^{2}+\omega ^{2}%
}{\omega ^{2}}\right) ^{\frac{1}{2}}$, $\gamma =-\arctan \left( \frac{\eta }{%
\omega }\right) $. As a result, the gaussian width describing the electron
in the lowest Landau level ($LLL$) state would get reduced. Furthermore the
current operator, lying in the Hall direction in the absence of dissipation,
for $\eta \neq 0$ would acquire a longitudinal component, with a resulting
metallic conductance $\sigma _{L}\neq 0$ for the multielectron system. It is
remarkable that such an effect was soon after proposed in Ref. \cite%
{maldacena} in the context of boundary conformal field theories (BCFT). The
authors consider a system of two massless free scalar fields which have a
boundary interaction with a periodic potential and furthermore are coupled
to each other through a boundary magnetic term, whose expressions coincide
with Eqs. (\ref{mb2}) and (\ref{mb3}). By using a string analogy, the
boundary magnetic interaction allows for exchange of momentum of the open
string moving in an external magnetic field. Indeed it enhances one
chirality with respect to the other producing the same effect of a rotation
together with a scale transformation on the fields as for the dissipative
system of Ref. \cite{cgm5}, where the string parameter plays the role of
dissipation. It is crucial to observe that conformal invariance of the
theory is preserved only at special values of the parameters entering the
action, the so called ``magic''\ points.

Our recent work \cite{AV1}\cite{AV2} aimed at building up a general
effective theory for QHF which could add further evidence to the
relationship between the string theory picture and the condensed matter
theory one as well as to the role of noncommutativity in QHF physics. We
focused on a particular conformal field theory (CFT), the one obtained via $%
m $-reduction technique, which has been successfully applied to the
description of a quantum Hall fluid (QHF) at Jain \cite{cgm1} as well as
paired states fillings \cite{cgm2}\cite{cgm4} and in the presence of
topological defects \cite{noi1}\cite{noi2}\cite{noi5}. In particular, we
showed by means of the Morita equivalence that a NCFT with $\theta =2p+\frac{%
1}{m}$ or $\theta =\frac{p}{2}+\frac{1}{m}$ respectively is mapped to a CFT
on an ordinary space. We identified such a CFT with the $m$-reduced CFT
developed for a QHF at Jain $\nu =\frac{m}{2pm+1}$ \cite{cgm1}, as well as
paired states fillings $\nu =\frac{m}{pm+2}$ \cite{cgm2}\cite{cgm4}, whose
neutral fields satisfy twisted boundary conditions. In this way we gave a
meaning to the concept of ''noncommutative conformal field theory'', as the
Morita equivalent version of a CFT defined on an ordinary space. The image
of Morita duality in the ordinary space is given by the $m$-reduction
technique and the corresponding noncommutative torus Lie algebra is
naturally realized in terms of Generalized Magnetic Translations (GMT). That
introduces a new relationship between noncommutative spaces and QHF and
paves the way for further investigations on the role of noncommutativity in
the physics of general strongly correlated many body systems \cite%
{ncmanybody}.

In this paper we follow such a line of research and concentrate on the
physics of QHF at non standard fillings and in the presence of impurities.
We focus on the $m=2$ case which is experimentally relevant and describes a
system of two parallel layers of $2D$ electrons gas in a strong
perpendicular magnetic field and interacting with an impurity placed
somewhere on the boundary. The $2$-reduced theory on the two-torus obtained
as a Morita dual starting from a NCFT keeps track of noncommutativity in its
structure. Indeed GMT are a realization of the noncommutative torus Lie
algebra. We analyze in detail the presence of an impurity placed between the
layers somewhere on the edges and briefly recall the relation between
different impurities and different possible boundary conditions, introduced
in Ref. \cite{noi1}. A boundary state can be defined in correspondence to
each class of defects and a boundary partition function can be computed
which corresponds to a boundary fixed point, e. g. to a different
topological sector of the theory on the torus. The aim of this paper is to
show that the GMT are identified with operators which act on the boundary
states and realize the transition between fixed points of the boundary flow.
In the language of Kondo effect \cite{affleck1} they behave as boundary
condition changing operators. In this work we construct the most general GMT
operators which reflect the structure of our CFT model. They are introduced
as a tensor product, acting on the QHF and defect space respectively. Then
their role as boundary condition changing operators is discussed in detail
and its connection with non-Abelian statistics of the quasi-hole excitations
fully elucidated. Finally, the emergence of noncommutativity as a
consequence of the presence of the topological defect is emphasized.

The paper is organized as follows.

In Section 2, we briefly review the description of a QHF at paired states
fillings $\nu =\frac{m}{pm+2}$ obtained by $m$-reduction \cite{cgm2}\cite%
{cgm4} focusing on the case $m=2$, which describes a quantum Hall bilayer.
Then we point out that such a theory, when defined on a two-torus, is the
Morita dual of a NCFT whose noncommutativity parameter $\theta $ is rational
and expressed in terms of the filling fraction of the corresponding QHF.
Furthermore a realization of noncommutative torus Lie algebra is given by
Generalized Magnetic Translations (GMT).

In Section 3, we recall the different possible boundary interactions of our
quantum Hall bilayer model and briefly show how it is equivalent to a system
of two massless scalar bosons with a magnetic boundary interaction as
introduced in \cite{maldacena}, at the ``magic''\ points. Finally we point
out analogies with a system of strings and $D$-branes in a background field.

In Section 4, we express the boundary content of our theory in terms of
boundary partition functions and show that they are closed under the action
of GMT. In particular, we exploit the role of GMT as boundary condition
changing operators. These results allow us to infer the structure of the
most general GMT operators.

In Section 5, we discuss in detail the GMT structure of our CFT model for
the quantum Hall bilayer at paired states fillings and clarify the deep
relation between noncommutativity and non-Abelian statistics of quasi-hole
excitations, as a consequence of the presence of an impurity.

In Section 6, some comments and outlooks of this work are given.

Finally, the operator content of our theory, the TM, on the torus is
recalled in the Appendix.

\section{$m$-reduction description of a QHF at paired states fillings and
Morita equivalence}

In this Section we review how the $m$-reduction procedure on the plane works
in describing successfully a QHF at paired states fillings $\nu =\frac{m}{%
pm+2}$ \cite{cgm2,cgm4}. We focus mainly on the special case $m=2$ and on
the physics of a quantum Hall bilayer. Then we put in evidence that, on a
two-torus, the $m$-reduced theory is the Morita dual of an abelian NCFT with
rational noncommutativity parameter $\theta $.

The idea is to build up an unifying theory for all the plateaux with even
denominator starting from the bosonic Laughlin filling $\nu =1/pm+2$, which
is described by a CFT (mother theory) with $c=1$, in terms of a scalar
chiral field compactified on a circle with radius $R^{2}=1/\nu =pm+2$ (or
the dual $R^{2}=4/pm+2$):
\begin{equation}
Q(z)=q-i\,p\,lnz+\sum_{n\neq 0}\frac{a_{n}}{n}z^{-n},  \label{modes}
\end{equation}
with $a_{n}$, $q$ and $p$ satisfying the commutation relations $\left[
a_{n},a_{n^{\prime }}\right] =n\delta _{n,n^{\prime }}$ and $\left[ q,p%
\right] =i$. The corresponding primary fields are expressed in terms of the
vertex operators $U^{\alpha }(z)=:e^{i\alpha Q(z)}:$ with $\alpha
^{2}=1,...,2+pm$ and conformal dimension $h=\frac{\alpha ^{2}}{2}$.

Starting with this set of fields and using the $m$-reduction procedure,
which consists in considering the subalgebra generated only by the modes in
Eq. (\ref{modes}), which are a multiple of an integer $m$, we get the image
of the twisted sector of a $c=m$ orbifold CFT (daughter theory), the twisted
model (TM), which describes the lowest Landau level dynamics. Then the
fields in the mother CFT can be factorized into irreducible orbits of the
discrete $Z_{m}$ group which is a symmetry of the daughter theory and can be
organized into components which have well defined transformation properties
under this group. The general characteristics of the daughter theory is the
presence of twisted boundary conditions (TBC) which are induced on the
component fields and are the signature of an interaction with a localized
topological defect \cite{noi1,noi2,noi5}.

Let us concentrate on the special $m=2$ case, which is of our interest in
this paper. The filling factor $\nu ^{(a)}=\frac{1}{2p+2}$ is the same for
the two $a=1$, $2$ layers while the total filling is $\nu =\nu ^{(1)}+\nu
^{(2)}=\frac{1}{p+1}$. The CFT description for such a system can be given in
terms of two compactified chiral bosons $Q^{(a)}$ with central charge $c=2$.
In order to construct the fields $Q^{(a)}$ for the TM, the starting point is
the bosonic filling $\nu =1/2(p+1)$, described by a CFT with $c=1$ in terms
of a compactified scalar chiral field $Q$, as given in Eq. (\ref{modes}).
The $m$-reduction procedure generates a daughter theory which is a $c=2$
orbifold. Its primary fields content can be expressed in terms of a $Z_{2}$%
-invariant scalar field $X(z)$, given by
\begin{equation}
X(z)=\frac{1}{2}\left( Q^{(1)}(z)+Q^{(2)}(-z)\right) ,  \label{X1}
\end{equation}
describing the electrically charged sector of the new filling, and a twisted
field
\begin{equation}
\phi (z)=\frac{1}{2}\left( Q^{(1)}(z)-Q^{(2)}(-z)\right) ,  \label{phi1}
\end{equation}
which satisfies the twisted boundary conditions $\phi (e^{i\pi }z)=-\phi (z)$
and describes the neutral sector \cite{cgm2}. Such TBC signal the presence
of a localized topological defect which couples the $2$ edges through a
crossing \cite{noi1}\cite{noi2}\cite{noi5}, as shown in Fig. 1.

\begin{figure}[th]
\centering\includegraphics*[width=0.8\linewidth]{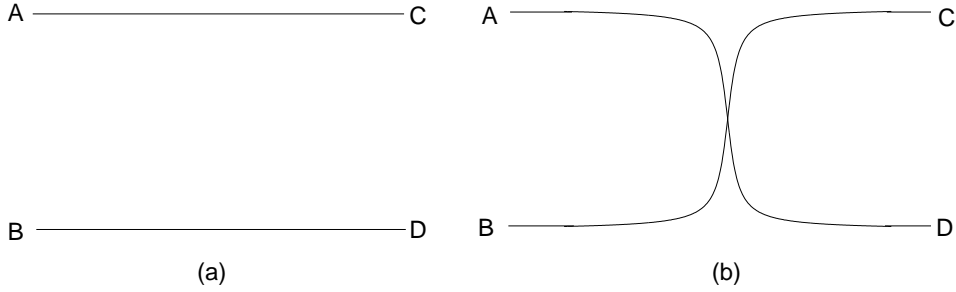}
\caption{The bilayer system, (a) without the topological defect (PBC), (b)
with the topological defect (TBC).}
\label{figura1}
\end{figure}

The primary fields are the composite operators $V(z)=\mathcal{U}_{X}(z)\psi
(z)$, where $\mathcal{U}_{X}(z)=\frac{1}{\sqrt{z}}:e^{i\alpha X(z)}:$ are
the vertices of the charged sector with $\alpha ^{2}=2(p+1)$. Furthermore
the highest weight states of the neutral sector can be classified in terms
of two kinds of chiral operators, $\psi (z)\left( \bar{\psi}(z)\right) =%
\frac{1}{2\sqrt{z}}\left( e^{i\alpha {\cdot }\phi (z)}\pm ie^{i\alpha {\cdot
}\phi (-z)}\right) $, which, in a fermionic language, correspond to $c=1/2$
Majorana fermions with periodic (Ramond) or anti-periodic (Neveu-Schwarz)
boundary conditions \cite{cgm4}. As a consequence this theory decomposes
into a tensor product of two CFTs, a twisted invariant one with $c=3/2$,
realized by the charged boson $X(z)$ and the Ramond Majorana fermion,\ which
is coupled to the charged sector, while the second one has $c=1/2$ and is
realized in terms of the Neveu-Schwarz Majorana fermion. The two Majorana
fermions just defined are inequivalent, due to the breaking of the symmetry
which exchanges them and that results in a non-Abelian statistics. If we
refer to the bilayer system, we can reduce the spacing between the layers so
that the two species of electrons which live on them become
indistinguishable and a flow to the Moore-Read (MR) states \cite{MR} takes
place. Furthermore $m$-ality in the neutral sector is coupled to the charged
one exactly, according to the physical request of locality of the electrons
with respect to the edge excitations: our projection, when applied to a
local field, automatically couples the discrete $Z_{m}$ charge of $U(1)$
with the neutral sector in order to give rise to a single valued composite
field.

Let us put our $m$-reduced theory on a two-torus and consider its
noncommutative counterpart $\mathrm{T}_{\theta }^{2}$, where $\theta $ is
the noncommutativity parameter. As shown in a recent paper \cite{AV2}, the $%
m $-reduction technique applied to the QHF at paired states fillings ($\nu =%
\frac{m}{pm+2}$, $p$ even) can be viewed as the image of the Morita map \cite%
{morita}\cite{scw1}\cite{scw2} (characterized by $a=\frac{p}{2}\left(
m-1\right) +1$, $b=\frac{p}{2}$, $c=m-1$, $d=1$) between the two NCFTs with $%
\theta =1$ and $\theta =\frac{p}{2}+\frac{1}{m}$ ($\theta =\nu _{0}/\nu ,$
being $\nu _{0}=1/2$ the filling of the starting theory), respectively and
corresponds to the Morita map in the ordinary space. The $\theta =1$ theory
is an $U\left( 1\right) _{\theta =1}$ NCFT while the mother CFT is an
ordinary $U\left( 1\right) $ theory; furthermore, when the $U\left( 1\right)
_{\theta =\frac{p}{2}+\frac{1}{m}}$ NCFT is considered, its Morita dual CFT
has $U\left( m\right) $ symmetry. Summarizing, the following correspondence
Table between the NCFTs and the ordinary CFTs is established:
\begin{equation}
\begin{array}{ccc}
& \text{Morita} &  \\
U\left( 1\right) _{\theta =1} & \rightarrow & U\left( 1\right) _{\theta =0}
\\
& \left( a=1,b=-1,c=0,d=1\right) &  \\
\text{Morita}\downarrow \left( a,b,c,d\right) &  & m-\text{reduction}%
\downarrow \\
& \text{Morita} &  \\
U\left( 1\right) _{\theta =\frac{p}{2}+\frac{1}{m}} & \rightarrow & U\left(
m\right) _{\theta =0} \\
& \left( a=m,b=-\frac{pm}{2}-1,c=1-m,d=\frac{p}{2}\left( m-1\right) +1\right)
&
\end{array}
\label{morita2}
\end{equation}%
Let us notice that theories which differ by an integer in the
noncommutativity parameter are not identical because they differ from the
point of view of the CFT. In fact, the Morita map acts on more than one
parameter of the theory.

We identify two kinds of noncommutativity related to the components of the
full magnetic field. The first one is due to the $B_{\parallel }$ component,
which defines the fillings of the $m$ layers and the second one is due to
the magnetic field $B_{\perp }$\ perpendicular to the layers. The
noncommutativity discussed in this paper is related only to the $B_{\perp }$
component due to its action on the defects (i.e. boundary magnetic fields
for the quantum Hall fluid CFT), though the $B_{\parallel }$ one also gives
important effects on the theory structure.

The compactification radius of the charged component is renormalized to $%
R_{X}^{2}=p+\frac{2}{m}$,\ that gives rise to different CFTs by varying $p$
values. Moreover the action of the $m$-reduction procedure on the number $p$
doesn't change the central charge of the CFT under study but modifies the
spectrum of the charged sector \cite{cgm2}\cite{cgm4}. Furthermore the
twisted boundary conditions on the neutral fields of the $m$-reduced theory,
Eq. (\ref{phi1})), arise as a consequence of the noncommutative nature of
the $U\left( 1\right) _{\theta =\frac{p}{2}+\frac{1}{m}}$ NCFT.

The key role in the proof of equivalence is played by the map on the field $%
Q(z)$\ of Eq. (\ref{modes}) which, after the Morita action, is defined on
the noncommutative space $z\rightarrow z^{1/m}\equiv U_{0,1}$. The
noncommutative torus Lie algebra, defined by the following commutation
rules:
\begin{equation}
\left[ U_{\overrightarrow{n}+\overrightarrow{j}},U_{\overrightarrow{%
n^{\prime }}+\overrightarrow{j^{\prime }}}\right] =-2i\sin \left( 2\pi
\theta \overrightarrow{j}\wedge \overrightarrow{j^{\prime }}\right) U_{%
\overrightarrow{n}+\overrightarrow{n^{\prime }}+\overrightarrow{j}+%
\overrightarrow{j^{\prime }}},  \label{fexp5}
\end{equation}
is realized in terms of the $m^{2}-1$ general operators:
\begin{equation}
\begin{array}{cc}
U_{j_{1},j_{2}}=\varepsilon ^{\frac{j_{1}j_{2}}{2}}z^{j_{1}}\varepsilon
^{j_{2}\widetilde{\sigma }}, &
\begin{array}{c}
j_{1},j_{2}=0,...,m-1 \\
\left( j_{1},j_{2}\right) \neq \left( 0,0\right)%
\end{array}%
\end{array}
,  \label{circles4}
\end{equation}
\bigskip where $\widetilde{\sigma }=iz\partial _{z}$. Via Morita duality, a
mapping between a general field operator $\mathbf{\Phi }$ defined on the
noncommutative torus \textrm{T}$_{\theta }^{2}$ and the field $\Phi $ living
on the dual commutative torus $\mathrm{T}_{\theta =0}^{2}$ is generated as
follows:
\begin{equation}
\mathbf{\Phi }=\sum_{\overrightarrow{n}}\exp \left( 2\pi im\frac{%
\overrightarrow{n}\cdot \widehat{\overrightarrow{x}}}{R}\right) \sum_{%
\overrightarrow{j}=0}^{m-1}\Phi ^{\overrightarrow{n},\overrightarrow{j}}U_{%
\overrightarrow{n}+\overrightarrow{j}}\longleftrightarrow \Phi =\sum_{%
\overrightarrow{j}=0}^{m-1}\chi ^{\left( \overrightarrow{j}\right) }J_{%
\overrightarrow{j}}.  \label{fexp9}
\end{equation}
The new field $\Phi $ is defined on the dual torus with radius $R^{^{\prime
}}=\frac{R}{m}$ and satisfies the \textit{twist eaters} boundary conditions:
\begin{equation}
\begin{array}{cc}
\Phi \left( x_{1}+R^{\prime },x_{2}\right) =\Omega _{1}^{+}\cdot \Phi \left(
x_{1},x_{2}\right) \cdot \Omega _{1}, & \Phi \left( x_{1},x_{2}+R^{\prime
}\right) =\Omega _{2}^{+}\cdot \Phi \left( x_{1},x_{2}\right) \cdot \Omega
_{2},%
\end{array}
\label{fexp12}
\end{equation}
with
\begin{equation}
\Omega _{1}=\mathrm{P}^{c},\text{ \ \ \ \ \ \ }\Omega _{2}=\mathrm{Q},
\label{fexp13}
\end{equation}
where $c$ is the 't Hooft magnetic flux, $\mathrm{P}$ and $\mathrm{Q}$ are
the so called ''shift''\ and ''clock''\ generators \cite{matrix2}\cite%
{matrix3} and the field components $\chi ^{\left( \overrightarrow{j}\right)
} $ satisfy twisted boundary conditions.

By using the above decomposition (Eq. (\ref{fexp9})), where $\Phi \equiv
Q(z) $, we identify the fields $X(z)$\ and $\phi ^{j}(z)$\ of the CFT
defined on the ordinary space. Indeed $\chi ^{\left( 0,0\right) }$ is the
trace degree of freedom which can be identified with the $U(1)$ component of
the matrix valued field or the charged $X$ field (\ref{X1}) within the $m$%
-reduced theory of the QHF, while the twisted fields $\chi ^{\left(
\overrightarrow{j}\right) }$ with $\overrightarrow{j}\neq \left( 0,0\right) $
should be identified with the neutral ones (\ref{phi1}).

In conclusion, when the parameter $\theta $ is rational we recover the whole
structure of the noncommutative torus and recognize the twisted boundary
conditions which characterize the neutral fields (\ref{phi1}) of the $m$%
-reduced theory as the consequence of the Morita mapping of the starting
NCFT ($U\left( 1\right) _{\theta =\frac{p}{2}+\frac{1}{m}}$ in our case) on
the ordinary commutative space. In such a picture the GMT are a realization
of the noncommutative torus Lie algebra defined in Eq. (\ref{fexp5}).

In the following Section we introduce the different possible boundary
conditions for the quantum Hall bilayer system under study and write down
the corresponding boundary action. Then the equivalence with a system of two
massless scalar bosons with a magnetic boundary interaction \cite{maldacena}
at the ``magic''\ points is recalled \cite{noi2}. Finally, an explicit
correspondence with a system of strings and $D$-branes in a background $B$%
-field is traced.

\section{$m$-reduced CFT for QHF at paired states fillings and boundary
interactions}

In this Section we summarize the different possible boundary conditions of
our CFT model for the quantum Hall bilayer system and point out its
equivalence with a system of two massless scalar bosons with a magnetic
boundary interaction at ``magic''\ points \cite{noi1}\cite{noi2}\cite{noi5}.
Finally we outline the correspondence with a system of strings and $D$%
-branes in a background $B$-field.

Our TM theory is the continuum description of the quantum Hall bilayer under
study. Its key feature is the presence of two different boundary conditions
for the fields defined on the two layers:
\begin{equation}
\varphi _{L}^{\left( 1\right) }\left( x=0\right) =\pm \varphi _{R}^{\left(
2\right) }\left( x=0\right) ,  \label{blr}
\end{equation}
where the $+$ ($-$) sign identifies periodic (PBC) and twisted (TBC)
boundary conditions respectively, $L$ and $R$ staying for left and right
components. Indeed TBC are naturally satisfied by the twisted field $\phi
\left( z\right) $ of our TM (see Eq. (\ref{phi1})), which describes both the
left moving component $\varphi _{L}^{\left( 1\right) }$ and the right moving
one $\varphi _{R}^{\left( 2\right) }$ in a folded description of a system
with boundary. In the limit of strong coupling they account for the
interaction between a topological defect at the point $x=0$ (i. e. the
layers crossing shown in Fig. 1) and the up and down edges of the bilayer
system. When going to the torus topology, the characters of the theory are
in one to one correspondence with the ground states and a doubling of the
corresponding degeneracy is expected, which can be seen at the level of the
conformal blocks (see Appendix). Indeed we get for the PBC case an untwisted
sector, $P-P$ and $P-A$, described by the conformal blocks (\ref{vacuum1})-(%
\ref{ut3}), and for the TBC case a twisted sector, $A-P$ and $A-A$,
described by the conformal blocks (\ref{tw1})-(\ref{tw6}). Summarizing, the
two layer edges can be disconnected or connected in different ways, implying
different boundary conditions, which can be discussed referring to the
characters with the implicit relation to the different boundary states (BS)
present in the system (see Ref. \cite{noi1}). These BS should be associated
to different kinds of linear defects compatible with conformal invariance
and their relative stability can be established. The knowledge of the
relative stability of the different boundary states is crucial for the
reconstruction of the whole boundary renormalization group (RG) flow. Indeed
different boundary conditions correspond to different classes of boundary
states, each one characterized by a $g$-function \cite{Affleck2}, the $g$%
-function decreases along the RG flow when going form the UV to the IR fixed
point \cite{noi1} and the generalized magnetic translations play the role of
boundary condition changing operators, as we will show in the following
Sections.

Let us now write down the action for our bilayer system in correspondence of
the different boundary conditions imposed upon it, i. e. PBC and TBC. In the
absence of an edge crossing (PBC case) the Hamiltonian of the bilayer system
is simply:
\begin{equation}
H=\frac{1}{2}\left[ \left( \Pi ^{\left( 1\right) }\right) ^{2}+\left( \Pi
^{\left( 2\right) }\right) ^{2}+\left( \partial _{x}Q^{(1)}\right)
^{2}+\left( \partial _{x}Q^{(2)}\right) ^{2}\right],  \label{hampbc}
\end{equation}
where $Q^{(1)}$ and $Q^{(2)}$ are the two boson fields generated by $2$%
-reduction and defined on the layers $1$ and $2$ respectively (see Section
2), while the presence of such a coupling (TBC case, see Fig. 1) introduces
a magnetic twist term of the kind:
\begin{equation}
H_{M}=\beta \left( Q^{(1)}\partial _{t}Q^{(2)}-Q^{(2)}\partial
_{t}Q^{(1)}\right) \delta \left( x\right) .  \label{magterm}
\end{equation}
Finally, in the presence of a localized defect (or a quantum point contact)
the Hamiltonian contains a boundary tunneling term such as:
\begin{equation}
H_{P}=-t_{P}\cos \left( Q^{(1)}-Q^{(2)}\right) \delta \left( x\right) ,
\label{pot1}
\end{equation}
which implements a locally applied gate voltage $V_{g}=t_{P}\delta \left(
x\right) $. Thus the full Hamiltonian can be written as \cite{noi2}\cite%
{noi5}:
\begin{eqnarray}
H &=&\frac{1}{2}\left[ \left( \Pi ^{\left( 1\right) }\right) ^{2}+\left( \Pi
^{\left( 2\right) }\right) ^{2}+\left( \partial _{x}Q^{(1)}\right)
^{2}+\left( \partial _{x}Q^{(2)}\right) ^{2}\right] -t_{P}\cos \left(
Q^{(1)}-Q^{(2)}\right) \delta \left( x\right)  \notag \\
&&+\beta \left( Q^{(1)}\partial _{t}Q^{(2)}-Q^{(2)}\partial
_{t}Q^{(1)}\right) \delta \left( x\right) .
\end{eqnarray}
Notice that the boundary tunneling term is proportional to $\phi $ and the
magnetic term produces a twist on $\phi $.

Our bilayer system looks like very similar to a system of two massless
scalar fields $X$ and $Y$ \ in $1+1$ dimensions, which are free in the bulk
except for boundary interactions, which couple them. Its action is given by $%
S=S_{bulk}+S_{pot}+S_{mag}$ \cite{maldacena} where:
\begin{eqnarray}
S_{bulk} &=&\frac{\alpha }{4\pi }\int_{0}^{T}dt\int_{0}^{l}d\sigma \left(
\left( \partial _{\mu }X\right) ^{2}+\left( \partial _{\mu }Y\right)
^{2}\right) , \\
S_{pot} &=&\frac{V}{\pi }\int_{0}^{T}dt\left( \cos X\left( t,0\right) +\cos
Y\left( t,0\right) \right) ,  \label{SP} \\
S_{mag} &=&i\frac{\beta }{4\pi }\int_{0}^{T}dt\left( X\partial
_{t}Y-Y\partial _{t}X\right) _{\sigma =0}.  \label{SM}
\end{eqnarray}
Here $\alpha $ determines the strength of dissipation and is related to the
potential $V$ while $\beta $ is related to the strength of the magnetic
field $B$ orthogonal to the $X-Y$ plane, as $\beta =2\pi B$. The magnetic
term introduces a coupling between $X$ and $Y$ at the boundary while keeping
conformal invariance. Such a symmetry gets spoiled by the presence of the
interaction potential term except for the magic points $\left( \alpha ,\beta
\right) =\left( \frac{1}{n^{2}+1},\frac{n}{n^{2}+1}\right) ,n\in \mathbb{Z}$%
. For such parameters values the theory is conformal invariant for any
potential strength $V$. Furthermore, if $\alpha =\beta $ there is a complete
equivalence between our TM model for the bilayer system and the above
boundary CFT, as shown in Ref. \cite{noi2}. All the degrees of freedom of
such a system are expressed in terms of boundary states, which can be easily
constructed by considering the effect of the magnetic interaction term as
well as that of the potential term on the Neumann boundary state $|N>$. In
this way one obtains the generalized boundary state $|B_{V}>$ as:
\begin{equation}
|B_{V}>=\sec \left( \frac{\delta }{2}\right) e^{i\delta \mathcal{R}%
_{M}}e^{-H_{pot}\left( 2X_{L}^{^{\prime }}\right) -H_{pot}\left(
2Y_{L}^{^{\prime }}\right) }|N^{X^{^{\prime }}}>|N^{Y^{^{\prime }}}>,
\end{equation}
where the rotation operator $\mathcal{R}_{M}$ is given by
\begin{equation}
\mathcal{R}_{M}=(y_{L}^{0}p_{L}^{X}-x_{L}^{0}p_{L}^{Y})+\sum_{n>0}\frac{i}{n}%
\left( \alpha _{n}^{Y}\alpha _{-n}^{X}-\alpha _{-n}^{Y}\alpha _{n}^{X}\right)
\label{rot1}
\end{equation}
and the rotation parameter $\delta $ is defined in terms of the parameters $%
\alpha ,\beta $ as $\tan \left( \frac{\delta }{2}\right) =\frac{\beta }{%
\alpha }$. Furthermore, the rotated and rescaled coordinates $X^{^{\prime
}},Y^{^{\prime }}$ have been introduced as:
\begin{eqnarray}
X^{^{\prime }} &=&\cos \frac{\delta }{2}\left( \cos \frac{\delta }{2}X-\sin
\frac{\delta }{2}Y\right) ,  \notag \\
Y^{^{\prime }} &=&\cos \frac{\delta }{2}\left( \sin \frac{\delta }{2}X+\cos
\frac{\delta }{2}Y\right) .  \label{XY1}
\end{eqnarray}
Finally the boundary partition function $Z_{NB_{V}}$ can be computed as:
\begin{equation}
Z_{NB_{V}}=\sec \left( \frac{\delta }{2}\right) =<N|q^{L_{0}+\widetilde{L}%
_{0}}|B_{V}>  \label{part1}
\end{equation}
because, in the open string language, the rotation $\mathcal{R}_{M}$
introduces now twisted boundary conditions in the $\sigma $ direction. In
order to better clarify the equivalence of our theory with the above system
of two massless scalar bosons with boundary interaction at ``magic''\ points
it has been shown that the interlayer interaction is diagonalized by the
effective fields $X,\phi $ of Eqs. (\ref{X1}) and (\ref{phi1}), which are
related to the layers fields $Q^{(1)},Q^{(2)}$ just by the relation given in
Eq. (\ref{XY1}) for $\alpha =\beta $ \cite{noi2}. Indeed they can be
rewritten as:
\begin{eqnarray}
X(z) &=&\cos (\varphi /4)\left( \sin (\varphi /4)Q^{(1)}(z)+\cos (\varphi
/4)Q^{(2)}(z)\right) \\
\phi (z) &=&\cos (\varphi /4)\left( \cos (\varphi /4)Q^{(1)}(z)-\sin
(\varphi /4)Q^{(2)}(z)\right) .
\end{eqnarray}
Such a transformation consists of a scale transformation plus a rotation;
for $\varphi =\pi $ the fields $X(z)$ and $\phi (z)$ of Eqs. (\ref{X1}) and (%
\ref{phi1}) are obtained and the transformations above coincide with the
transformations given in Eqs. (\ref{XY1}) for $\delta =\frac{\pi }{2}$. In
this context the boundary state $|B_{0}(\delta )>$ for the (untwisted)
twisted sector in the folded theory is obtained from the rotation $\mathcal{R%
}_{M}$ on the Neumann boundary state $|N(\theta )>$ when $\delta =0,\frac{%
\pi }{2}$ respectively and can be seen as due to a boundary magnetic term
according to \cite{maldacena}.

Let us now briefly trace another correspondence which allows us to make
further contact with noncommutativity. Concerning the boundary magnetic term
(\ref{SM}), it allows for exchange of momentum of the open string moving in
an external magnetic field. Indeed, if we refer to a worldsheet field theory
for open strings attached to $D$-branes, it can be rewritten as in Eq. (\ref%
{mb2}); as such, the limit $g_{ij}\sim l_{s}^{4}\sim \varepsilon \rightarrow
0$ while the field $B_{ij}$ is fixed, together with the canonical
quantization of the open string endpoint coordinates $y^{i}\left( t\right) $%
, induces a noncommutative geometry on the $D$-brane worldvolume and the
effective low-energy field theory results a noncommutative one for the
massless open string modes. Within a QHF picture the open massless strings
correspond to the Hall fluid and the boundary states realize the boundary
conditions due to the presence of massive $D$-branes on which open strings
end. Furthermore they are dyons carrying electric charges as well as
magnetic fluxes and, then, are influenced by the boundary potential $V$ and
magnetic field $B$ living on the $D$-brane. Concerning the boundary
potential term (\ref{SP}) it could be identified with a tachyonic boundary
term such as in Eq. (\ref{mb3}), with a cosine tachyonic profile.

In the following Section, starting from the boundary content of our TM we
explicitly compute the GMT action in a simple way, in terms of the
periodicity of the Jacobi theta functions which enter the boundary partition
functions $Z_{NB_{V}}\left( \delta ,V\right) $ of our model. In particular
we show how the defect interaction parameters $(V,\delta )$ change upon GMT.
This behavior characterizes GMT as boundary condition changing operators.

\section{Generalized Magnetic Translations as boundary condition changing
operators}

The aim of this Section is to show that boundary partition functions $%
Z_{NB_{V}}(\delta ,V)$ \cite{noi2}, which express the boundary content of
our theory, are closed under the action of GMT. That will be proven by
studying how the defect interaction parameters $(V,\delta )$ change upon
GMT. As a result, we find that the translations of electrons as well as
anyons form a closed algebra (the supersymmetric sine algebra (SSA) given in
the next Section, Eqs. (\ref{ssa1})-(\ref{ssa3})), where the parameters $V$
and $\delta $ remain unchanged modulo $m$. Indeed the stability algebra for
a fixed point identified by a fractional value of $(V,\delta )$ \ is given
by this subalgebra and a subset of the conformal blocks for the boundary
partition function $Z_{NB_{V}}(\delta ,V)$. Also quasi-hole translations
form a closed algebra but the parameters $(V,\delta )$ will change together
with the corresponding boundary partition function. In this way the
transition to a different fixed point of the boundary flow is obtained or,
in other words, the switching between the untwisted and twisted vacua of our
TM is realized. As a result the role of GMT as boundary condition changing
operators clearly emerges.

In order to study the action of a GMT on the boundary partition functions $%
Z_{NB_{V}}(\delta ,V)$, let us start from the boundary content of our theory
in the simplest $m=2$, $p=0$ case, which corresponds to the first
non-trivial ``magic''\ point $\alpha =\beta =\frac{1}{2}$ in Ref. \cite%
{maldacena}. The action of the magnetic boundary term, (\ref{magterm}) or (%
\ref{SM}), on the Neumann state $|N>$ is obtained by defining a pair of
left-moving fermions as:
\begin{equation}
\begin{array}{ccc}
\psi _{1}=c_{1}e^{\frac{i}{2}\left( Q^{\left( 2\right) }+Q^{\left( 1\right)
}\right) }=c_{1}e^{iX}, &  & \psi _{2}=c_{2}e^{-\frac{i}{2}\left( Q^{\left(
2\right) }-Q^{\left( 1\right) }\right) }=c_{2}e^{i\phi },%
\end{array}%
\end{equation}
where $c_{i}$, $i=1,2$ are cocycles necessary for the anticommutation. By
splitting the two Dirac fermions into real and imaginary parts, $\varphi
_{1}=\psi _{11}+i\psi _{12}$, $\varphi _{2}=\psi _{21}+i\psi _{22}$, we get
four left-moving Majorana fermions given by $\psi =\left( \psi _{11},\psi
_{12},\psi _{21},\psi _{22}\right) =\left( \cos X,\sin X,\cos \phi ,\sin
\phi \right) $ and a corresponding set of right-moving ones. In this
language the magnetic boundary term acts only on the fourth Majorana fermion
as $\mathrm{R}_{M}=e^{2i\delta }$, where $\delta =0$ ($\delta =\frac{\pi }{2}
$) for the untwisted (twisted) sector of our theory, being its action the
identity for the other components, while the potential term acts on the
Majoranas as:
\begin{equation}
\mathrm{R}_{P}=\left(
\begin{array}{cccc}
\cos \left( 2V\right) & -\sin \left( 2V\right) & 0 & 0 \\
\sin \left( 2V\right) & \cos \left( 2V\right) & 0 & 0 \\
0 & 0 & \cos \left( 2V\right) & -\sin \left( 2V\right) \\
0 & 0 & \sin \left( 2V\right) & \cos \left( 2V\right)%
\end{array}
\right) .
\end{equation}
So, the overall rotation of the corresponding fermionic boundary states is $%
\mathrm{R}=\mathrm{R}_{M}\mathrm{R}_{P}$ and the partition function $Z_{AB}$
can be rewritten as:
\begin{equation}
Z_{NB_{V}}\left( \delta ,V\right) =\left\langle N\right| e^{-L\left( L_{0}+%
\bar{L}_{0}\right) }|B_{V}\left( \delta \right) >=\sqrt{2}\left( q\right)
^{-2/24}\prod_{n=1}^{\infty }\det \left( \mathrm{I}+q^{n-\frac{1}{2}}\mathrm{%
R}\right) ,
\end{equation}
where $|A>$ is the Neumann boundary state $|N>$, $|B>$ is the
magnetic-potential BS $|B_{V}>$, $q=e^{2i\pi \tau }$ and $\mathrm{I}$ is the
identity matrix. The final result is:
\begin{equation}
Z_{NB_{V}}\left( \delta ,V\right) =\sqrt{2}\left( \frac{\theta _{3}\left(
V|\tau \right) }{\eta \left( \tau \right) }\sqrt{\frac{\theta _{3}\left(
V|\tau \right) }{\eta \left( \tau \right) }}\right) \sqrt{\frac{\theta
_{3}\left( \delta +V|\tau \right) }{\eta \left( \tau \right) }},
\label{ZUV1}
\end{equation}
where $\delta =0$ ($\delta =\frac{\pi }{2}$) for the untwisted (twisted)
sector.

The value of the parameters $\delta $ and $V$ identifies a fixed point in
the boundary flow. Now, in order to compute the GMT action on these fixed
points and characterize GMT as boundary condition changing operators, let's
look at the transformation properties of the generalized Jacobi $\theta
_{e}\left( \omega |\tau \right) $ functions, $e=1,...,4$, for translations
along the cycles $A$ and $B$ of the two-torus. In particular, the
translations of interest for our study correspond to the following
transformations, $\omega \rightarrow \omega +a+b\tau $, of the $\omega $
parameter with $a,b=\frac{1}{2},1$. The result is:
\begin{equation}
\begin{tabular}{llll}
$\theta _{1}\left( \pi |\tau \right) =\theta _{1}\left( 0|\tau \right) ,$ & $%
\theta _{2}\left( \pi |\tau \right) =-\theta _{2}\left( 0|\tau \right) ,$ & $%
\theta _{3}\left( \pi |\tau \right) =\theta _{3}\left( 0|\tau \right) ,$ & $%
\theta _{4}\left( \pi |\tau \right) =\theta _{4}\left( 0|\tau \right) $ \\
$\theta _{1}\left( \pi \tau |\tau \right) =-q^{-1}\theta _{1}\left( 0|\tau
\right) ,$ & $\theta _{2}\left( \pi \tau |\tau \right) =q^{-1}\theta
_{2}\left( 0|\tau \right) ,$ & $\theta _{3}\left( \pi \tau |\tau \right)
=q^{-\frac{1}{2}}\theta _{3}\left( 0|\tau \right) ,$ & $\theta _{4}\left(
\pi \tau |\tau \right) =-q^{-\frac{1}{2}}\theta _{4}\left( 0|\tau \right) $
\\
$\theta _{1}\left( \frac{\pi }{2}|\tau \right) =\theta _{2}\left( 0|\tau
\right) ,$ & $\theta _{2}\left( \frac{\pi }{2}|\tau \right) =-\theta
_{1}\left( 0|\tau \right) ,$ & $\theta _{3}\left( \frac{\pi }{2}|\tau
\right) =\theta _{4}\left( 0|\tau \right) ,$ & $\theta _{4}\left( \frac{\pi
}{2}|\tau \right) =\theta _{3}\left( 0|\tau \right) $ \\
$\theta _{1}\left( \frac{\pi }{2}\tau |\tau \right) =iq^{-\frac{1}{8}}\theta
_{4}\left( 0|\tau \right) ,$ & $\theta _{2}\left( \frac{\pi }{2}\tau |\tau
\right) =q^{-\frac{1}{8}}\theta _{3}\left( 0|\tau \right) ,$ & $\theta
_{3}\left( \frac{\pi }{2}\tau |\tau \right) =q^{-\frac{1}{8}}\theta
_{2}\left( 0|\tau \right) ,$ & $\theta _{4}\left( \frac{\pi }{2}\tau |\tau
\right) =iq^{-\frac{1}{8}}\theta _{1}\left( 0|\tau \right) $%
\end{tabular}%
.  \label{table1}
\end{equation}

On the basis of these properties we obtain the action of GMT on the
parameters $V$ and $\delta $ which are related to the couplings in the
boundary interactions. In this way it is possible to sort out translations
that leave the vacua unchanged and translations that change boundary
conditions by making a transition from one vacuum to another. We find that
GMT factorize into two groups acting on $V$ and $\delta $ respectively as $V=%
\mathbf{\lambda }\frac{\pi }{2}=\frac{2\mathbf{l}+\mathbf{i}}{2}\pi $ and $%
\delta =\frac{\mathbf{i}}{2}\pi $. Only translations with $\mathbf{i}\neq 0$
(i. e. quasi-holes) change the boundary states, i. e. the fixed points
within the boundary flow, and then act as boundary condition changing
operators. For any fixed point we find a stability group which is the
subgroup of the GMT leaving the vacuum state unchanged: it is built of any
translation of particles with $\mathbf{i}=0$ (i. e. electrons and, for $%
p\neq 0$, anyons).

By taking a closer look to Eq. (\ref{ZUV1}) we clearly see that the
partition function for the MR model (for $p=0$) is given by the terms in the
bracket and depends only on the parameter $V$ wich is related to the
localized tunneling potential in Eqs. (\ref{SP}) or (\ref{pot1}). Both
charged and neutral components of the MR model translate together with the
same step as a result of the coupling between the two sectors due to the
parity rule. Let us notice that pure MR translations cannot be obtained
without the localized twist term, as shown in detail in the next Section. In
order to act on the MR states without modifying the last term in the
boundary partition function, Eq. (\ref{ZUV1}), it is mandatory to compensate
a $V$ translation with a translation in the $\delta $ parameter. This is a
consequence of the competition between the localized tunneling and the layer
exchange effects.

At this point we are ready to combine the action of GMT above obtained with
the modular properties of the conformal blocks (see Appendix); as a result
the parameters $V$ and $\delta $ transform as elements of the group $%
SL\left( 2,Z\right) /\Gamma _{2}$, in agreement with the conjecture about
magic points by Callan et al. \cite{maldacena}.

The results just obtained lead us to construct general GMT operators which
embody the peculiar features of our model for the quantum Hall bilayer at
paired states fillings. Thus, in the following Section we discuss in detail
such a new GMT structure and focus on the relation between noncommutativity
and non-Abelian statistics of quasi-hole excitations.

\section{Noncommutativity and Generalized Magnetic Translations}

In this Section we present in detail for the first time the rich structure
of Generalized Magnetic Translations within our TM for the quantum Hall
bilayer at paired states fillings. This study is mandatory in order to sort
out noncommutativity and clarify its relation with non-Abelian statistics of
quasi-hole excitations.

Our aim is to show that a signature of noncommutativity within our TM can be
found on GMT that become non-Abelian. In fact the quantum numbers which
label the states do not satisfy simple additive rules of composition as in
the Laughlin series but a more complex rule similar to that of a spin. In
general GMT do not commute with the full chiral algebra but only with the
Virasoro one. Therefore a spectrum can be found for any of the different
vacua which correspond to different defects. In order to understand this
phenomenon, let us study in detail our TM model with its boundary structure.
We will realize that noncommutativity is deeply related to the presence of
topological defects.

The whole TM can be written as $U(1)\times PF_{2}^{m}\times PF_{m}^{2}$,
where $U(1)$ is the Abelian charge/flux sector and the remaining factors
refer to neutral sector. Indeed the neutral fields can be decomposed into
two independent groups: one realizes the parafermions of the $SU(2)_{m}$
affine algebra ($PF_{m}^{2}$) and gives rise to a singlet of the twist
algebra while the second one realizes the parafermions of the $SU(m)_{2}$
algebra ($PF_{2}^{m}$) and is an irrep of the twist group. In the literature
the neutral sector was assumed to be insensible to magnetic translations due
to the neutrality \cite{MR}. Nevertheless, we will show that this is not
true in general because the breaking of the $U(1)$ pseudospin group to a
discrete subgroup implies that the residual action of the magnetic
translation group survives. Noncommutativity in the MR subtheory $U(1)\times
PF_{2}^{m}$ of our TM arises as a result of the coupling between the charged
and the neutral component due to the $m$-ality selection rule (i. e. the
pairing phenomenon for $m=2$). In order to describe the antisymmetric part
of the full TM, instead, the coset $SU(m)_{2}/U(1)^{m-1}$must be taken into
account. The boundary interactions, Eqs. (\ref{magterm}) and (\ref{pot1}),
break the symmetry because they contain operators with pseudospin $\Sigma
^{2}$ giving rise to the different fixed points of the boundary phase
diagram (with a $\Sigma _{z}$ residual symmetry) \cite{noi1}. The twist
fields are the image of noncommutativity and the fixed points correspond to
different representations with different pseudospin. However, thanks to the
modular covariance \cite{cgm4}, we can go from a representation to the other.

\ Let us now study in detail the structure of GMT. The MR conformal blocks $%
\chi _{(\lambda ,s)}^{MR}$ (see Eqs. (\ref{mr1})-(\ref{mr3}) in the
Appendix) depend on the Abelian index $s=0,...,p$ and the spin index $%
\lambda =0,..,m$. The subgroup of GMT which stabilize the MR conformal
blocks is realized by the operators which transport symmetric electrons, as
we will show in the following.

The $U(1)$ Abelian charge/flux sector is characterized by a definite $%
(q,\phi )$-charge and flux for any type of particles, which simply add
together due to the charge and flux conservation. A phase $e^{iq\phi }$ is
generated by winding one particle around another (i.e. the Abelian
Aharonov-Bohm phase factor). In the presence of the neutral component this
phase is fairly simple. We can have a different behaviour depending on the
Abelian or non-Abelian nature of the excitations. In the untwisted sector
(i.e. without $\sigma $-fields) the particles, which are anyons or
electrons, all exhibit Abelian statistics. Nevertheless the neutral
components give a contribution to the full Aharonov-Bohm phase by means of
the $Z_{m}$ charge. For the simplest $m=2$ case (quantum Hall bilayer) of
our interest in this work we encode this charge into the fermion number $F$,
counting the fermion modes, which is defined by means of $\gamma _{F}$.
Notice that in our construction we do not need to introduce cocycles for
neutral modes because the induction procedure automatically gives the
correct commutation relations for the projected fields. Nevertheless, when
we consider the charged and neutral sectors as independent it is necessary
to consider such matrices. Thus, when we decompose the $c=1$ neutral modes
into two $c=1/2$ components (see Section 2 and Appendix) two independent
Clifford algebras for fermion zero modes have to be introduced. The twisted $%
|\sigma /\mu >$ ground state is degenerate and it is possible to define a
Clifford algebra in terms of Pauli matrices, which act as $\mathbf{\Sigma }%
_{x}|\sigma /\mu >=|\mu /\sigma >;$ $\mathbf{\Sigma }_{y}|\sigma /\mu >={\pm
}i|\mu /\sigma >;$ $\ \ \mathbf{\Sigma }_{z}|\sigma /\mu >={\pm }|\sigma
/\mu >$, and of the operator $\gamma _{F}=(-1)^{F}$, which is defined in
such a way to anticommute with the fermion field, $\gamma _{F}\psi \gamma
_{F}=-\psi $, and to satisfy the property $\left( \gamma _{F}\right) ^{2}=1$%
; furthermore it has eigenvalues $\pm 1$ when acting on states with even or
odd numbers of fermion creation operators. On the above vacua the Clifford
algebra is realized in terms of the fermion modes by means of the following
operators:
\begin{eqnarray}
\gamma _{F} &=&e^{i\frac{\pi }{4}\Sigma _{z}}(-1)^{\sum \psi _{-n}\psi _{n}}%
\text{ \ \ \ and \ \ }\psi _{0}=\frac{e^{i\frac{\pi }{4}\Sigma _{x}}}{\sqrt{2%
}}(-1)^{\sum \psi _{-n}\psi _{n}}\text{\ ,\ \ twisted vacuum}
\label{clifford} \\
\gamma _{F} &=&I(-1)^{\sum \psi _{-n+1/2}\psi _{n+1/2}}\text{ ,\ \ \ \ \ \ \
\ \ \ \ \ \ \ \ \ \ \ \ \ \ \ \ \ \ \ \ \ \ \ \ \ \ \ \ \ \ \ \ \ \ \ \ \ \
\ \ \ \ \ \ untwisted vacuum}
\end{eqnarray}
in a\ $\gamma _{F}$ diagonal basis. In order to give a unified
representation of the $\gamma _{F}$ operator in the twisted as well as the
untwisted space we need to add the identity operator $I$\ in the above
definition. It acts on the layer indices space (i. e. the pseudospin space).

Within the MR model there isn't a well defined $\gamma _{F}$ in the twisted
ground state, thus we need to take vacuum states of the form $|\tilde{\sigma}%
>_{\pm }=\frac{1}{\sqrt{2}}\left( |\sigma >\pm |\mu >\right) $ while modular
invariance forces us to consider only one of these states, which corresponds
to the $\chi _{\frac{1}{16}}$ character appearing in Eq. (\ref{mr2}) (see
Appendix). In terms of fields, this would mean trading the two fields $%
\sigma $ and $\mu $ for a single field $\tilde{\sigma}_{\pm }=\frac{1}{\sqrt{%
2}}\left( \sigma \pm \mu \right) $. The fusion rule $\psi \times \sigma =\mu
$ would be replaced by $\psi \times \tilde{\sigma}_{\pm }=\pm \tilde{\sigma}%
_{\pm }$. The MR\ model contains only one of these operators ($\tilde{\sigma}%
_{+}$) which corresponds to the $\chi _{\frac{1}{16}}$ character while the
characters $\chi _{0}$ and $\chi _{\frac{1}{2}}$ have a well defined fermion
parity. As it was observed on the plane (see Section 2), the charged and the
neutral sector of MR model are not completely independent but need to
satisfy the constraint $\lambda =\alpha \cdot p+l=0$ ($mod$ $2$) which is
the $m$-ality condition (parity rule). Here such a rule is explicitly
realized by constraining the eigenvalues of the fermion parity operator upon
defining the generalized GSO projector $P=\frac{1}{2}(1-e^{i\pi \alpha \cdot
p}\gamma _{F})$. \ In this way the eigenvalues of the integer part of the
neutral translation can be related to the charged one. In order to formally
extend the definition of magnetic translations within the neutral sector to
the transport of an Abelian anyon in any of the different vacua we introduce
the couple ($a,F$) of parameters which are defined only modulo $2$.
According to the above definition $a$ is equal to $1$ for twisted ($|\tilde{%
\sigma}>_{+}$) and $0$ for untwisted vacua ($|I>$ and $|\psi >$)
consistently with the fusion rules. Transport of an anyon around another one
produces a $(-1)^{a_{1}F_{2}-a_{2}F_{1}}$ phase and the following
identification holds: $(a,F)=(\lambda -i-2l,l+\frac{i+\lambda }{2})$, which
corresponds to the characteristics of the Jacobi theta functions $\theta %
\left[
\begin{array}{c}
\frac{a}{2} \\
\frac{F}{2}%
\end{array}%
\right] $ within Ising characters. In conclusion, for the MR sector the
parity rule coupling between the charged and neutral sector manifests itself
as the completion of the charge/flux quantum numbers $(q,a;\phi ,F)$.

On the basis of the previous considerations, the most general GMT operators
can be written as the tensor product $\mathcal{J}_{F}^{a}\otimes \mathcal{J}%
_{D}^{a}$, acting on the quantum Hall fluid and defects space respectively.
Although electrons and anyons do not exhibit non-Abelian statistics, the GMT
are different for the untwisted/twisted sectors as a consequence of the
difference in the definition of $\gamma _{F}$, which is a simple identity
operator in the untwisted space but becomes a spin operator in the twisted
one. It is easy to verify that the action of \ $\mathbf{\Sigma }_{x}$ on $%
\psi _{0}$ is simply the layer exchange which takes place when the fermion\
crosses\ the defect line. When a particle encyrcles a defect, it takes a
phase $(-1)^{F}$ and changes/unchanges the pseudospin depending on the $%
\mathbf{\Sigma }/I$ operator action. The net effect of the pseudospin is to
modify the GMT bracket into an anticommutator so that the GMT algebra
becomes, for these Abelian particles, a graded algebra. We can define two
kinds of generators:
\begin{eqnarray}
\mathcal{J}_{-}^{\mathbf{a}} &=&\mathcal{J}^{\mathbf{a}}\otimes \mathbf{I} \\
\mathcal{J}_{+}^{\mathbf{a}} &=&\mathcal{J}^{\mathbf{a}}\otimes \mathbf{%
\Sigma }_{z}
\end{eqnarray}%
(where $+/-$ refer to untwisted/twisted vacuum).

Here we note that the pseudospin operator which appears in $\mathcal{J}_{+}^{%
\mathbf{a}}$ is a direct consequence of the noncommutative structure of the
defect obtained in \cite{AV2}. It is made explicitly in terms of magnetic
translations operators $\mathcal{J}_{D}^{a}$ = $U_{j_{1},j_{2}}$ given in
Eq.(\ref{circles4}), which generate an algebra isomorphic to $SU(2)$.

A straightforward calculation tells us that $\mathcal{J}_{-}^{\mathbf{a}}$
and $\mathcal{J}_{+}^{\mathbf{a}}$ satisfy the following super-magnetic
translation algebra (within the MR sector):
\begin{eqnarray}
\left[ \mathcal{J}_{-}^{\mathbf{a}},\mathcal{J}_{-}^{\mathbf{\beta }}\right]
&=&2i\sin \left( \frac{\mathbf{s}{\times }\mathbf{s}^{\prime }}{p+1}\pi
\right) \mathcal{J}_{-}^{\mathbf{a}+\mathbf{\beta }},  \label{ssa1} \\
\left[ \mathcal{J}_{+}^{\mathbf{a}},\mathcal{J}_{-}^{\mathbf{\beta }}\right]
&=&2i\sin \left( \frac{\mathbf{s}{\times }\mathbf{s}^{\prime }}{p+1}\pi
\right) \mathcal{J}_{+}^{\mathbf{a}+\mathbf{\beta }},  \label{ssa2} \\
\left\{ \mathcal{J}_{+}^{\mathbf{a}},\mathcal{J}_{+}^{\mathbf{\beta }%
}\right\} &=&2\cos \left( \frac{\mathbf{s}{\times }\mathbf{s}^{\prime }}{p+1}%
\pi \right) \mathcal{J}_{-}^{\mathbf{a}+\mathbf{\beta }},  \label{ssa3}
\end{eqnarray}%
that is the supersymmetric sine algebra (SSA).

In order to clarify the deep relationship between noncommutativity and
non-Abelian statistics in our TM let us focus on $\mathcal{J}_{D}^{a}$
operators, noncommutativity being related to the presence of topological
defects on the edge of the quantum Hall bilayer. Only on the twisted vacuum
(i.e. for $V, \delta =\pi /2$) the $\mathcal{J}_{D}^{a}$\ realize the spin
operator $\mathbf{\Sigma }$ while, in the usual untwisted vacuum with $V,
\delta =\pi $, there is no noncommutativity and $\mathcal{J}_{D}^{a}$\
reduces to the identity $\mathbf{I}$. The defects break the GMT symmetry so
that different backgrounds can be connected by special GMT. The breaking of
the residual symmetry of the CFT can be recognized in the condensation of
the defects. As recalled in Section 3, it is possible to identify three
different classes of defects and then three non trivial fixed points; in
particular an intermediate coupling fixed point has been found \cite{noi1},
which could be identified with the non-Fermi liquid fixed point which
characterizes the overscreened two-channel Kondo problem in a quantum
impurity context \cite{affleck1}. The $\mathcal{J}_{D}^{a}$ operators which
act on the defects space are realized by means of the $\mathrm{P}$, $\mathrm{%
Q}$ operators defined on the noncommutative torus (see Eq. (\ref{fexp13}) in
Section 2 and Ref. \cite{AV2} for details) and are taken to build up the $%
SU(m)$ boundary generators. The Clifford algebra can be realized in terms of
these operators. The non-Abelian statistics is obtained in the presence of a
$\sigma /\mu $-twist that corresponds to the defects. A tunneling phenomenon
is associated to a twist of the MR sector while a level crossing is obtained
in the presence of a twist of the pure Ising sector. The \textquotedblright
boundary\textquotedblright\ $SU(2)$ algebra acts on the twisted boundary
conditions of the neutral fermions. The noncommutative nature arises in the
TM as a manifestation of the vacuum degeneracy of the non-Fermi liquid fixed
point, and the interaction with the defect spin (pseudospin) is given by the
Clifford algebra. Notice that this noncommutativity is purely chiral and
should not be confused with the Aharonov-Bohm effect due to charge/flux
exchange (which is necessarily non-chiral).

Let us now define another GMT algebra which is a subalgebra of the whole GMT
algebra; it is generated by the exchange of quasi-hole excitations with
non-Abelian statistics. The GMT operators for quasi-holes are more involved
and, as shown in the previous Section, change boundary states by switching
from a twisted vacuum to an untwisted one and viceversa; as a result a
parity operator $\Omega $ with the property $\Omega ^{2}=(-1)^{F}$ must be
inserted in the definition of generators producing $\mathcal{J}%
_{D}^{a}=\Omega e^{i\frac{\pi }{4}\Sigma }$. That corresponds to the
introduction of the phase factor $e^{-i\frac{\pi }{4}}$when considering the
action on the conformal blocks \cite{AV3}. Although only closed edges are
physical, this forces us to introduce open edges as the fundamental domain
of our theory. In a string theory context the operator $\Omega $ realizes
the switching from the closed string channel to the open string one ending
on a massive $D$-brane, identified with the topological defect. Furthermore
let us notice that the defects support Majorana fermion zero modes: this
finding in our context of QHF physics parallels an analogous recent finding
by Teo and Kane in topological insulators and superconductors \cite{teo1}.
It is well known \cite{nayak} that non-Abelian statistics of quasi-hole
excitations has a $SO(2n)$ structure, typical of the so called Ising anyons.
We can realize such an algebra in our formalism by using the non-diagonal
GMT. In fact, while the product of $2n$ identical one-particle translations
are a realization of the GMT on the $2n$-particles wave functions, the
product of two independent one-particle translations can be used to realize
the braiding matrices embedded in $SO(2n)$. Let us consider here only the
neutral translations and the tensor product of $2n$ copies of such
translations. In terms of field theory this corresponds to an Ising$^{2n}$
model. Indeed it is a standard realization of $SO(2n)$ algebra which groups
the $2n$ Majorana fields into $n$ complex Dirac fields. If also one-particle
translations are added, a superextension of this algebra is obtained \cite%
{isinganyons}. A representation of the braid group for $2n$ quasiholes has
dimension $2^{n-1}$and can be described as a subspace of the tensor product
of $2n$ two dimensional spaces \cite{nayak}. Each of them contains basis
vectors and the physical subspace of the tensor product is the space
generated by the vectors whose overall sign is positive. A spinor
representation of $SO(2n)\times U(1)$ lives on the tensor product space: the
$U(1)$ factor acts as a multiplicative factor, while the generators $\Sigma
_{ij}$ of $SO(2n)$ may be written in terms of the Pauli matrices $\Sigma
_{i} $. In conclusion, the braid group $\mathcal{B}_{n}$ is generated by
elementary exchanges $T_{i}$ of a quasihole $i$ and a quasihole $i+1$,
satisfying the relations:
\begin{eqnarray}
T_{i}T_{j} &=&T_{j}T_{i}\text{ \ \ \ \ \ \ \ (}|i-j|\geqslant 2\text{)}
\label{braid1} \\
T_{i}T_{i+1}T_{i} &=&T_{i+1}T_{i}T_{i+1}\text{ \ \ \ (}1\leq i\leq n-2\text{%
).}  \label{braid2}
\end{eqnarray}
They can be realized as embedded in the action of $SO(2n)\times U(1)$ as
follows: $T_{i}=\Omega e^{i\frac{\pi }{2}\Sigma _{ij}}$. The odd operators $%
T_{i}=\Omega e^{i\frac{\pi }{4}\Sigma _{i}}$ act as one-particle ones and
are identified as the GMT for a quasi-hole while the even ones, $%
T_{i}=\Omega e^{i\frac{\pi }{4}\Sigma _{i}\Sigma _{i+1}}$, are two-particle
operators. We can see that the square of a quasi-hole translation coincides
with the GMT for an electron. Thus we obtain the GMT group as the double
covering of the one-particle operators for the quasi-hole transport.

The action of diagonal GMT on the conformal blocks (i. e. the one-point
functions) of our TM, which is the relevant one for a study of the boundary
fixed points, will be the subject of a forthcoming publication \cite{AV3}.

\section{Conclusions and outlooks}

In this work we studied some aspects of the physics of QHF at paired states
fillings in a more general context, that of a NCFT. Indeed, when the
underlying $m$-reduced CFT is put on a two-torus it appears as the Morita
dual of an abelian NCFT. By analyzing the boundary interaction terms present
in the action we recognized a boundary magnetic term and a boundary
potential which, within a string theory picture, could describe an analogue
system of open strings with endpoints finishing on $D$-branes in the
presence of a background $B$-field and a tachyonic potential.

We focused on a quantum Hall bilayer in the presence of a localized impurity
(i. e. a topological defect) and briefly recalled the boundary state
structure corresponding to two different boundary conditions, the periodic
as well as the twisted boundary conditions respectively, which give rise to
different topological sectors on the torus \cite{noi1}\cite{noi2}\cite{noi5}%
. In this context the action of GMT operators on the boundary partition
functions has been computed and their role as boundary condition changing
operators fully evidenced. From such results we inferred the general
structure of GMT in our model, which we study in great detail, and clarified
the deep relation between noncommutativity and non-Abelian statistics of
quasi-hole excitations. Non-Abelian statistics of quasi-holes is crucial for
physical implementations of topological quantum computing in QHF systems
\cite{sarma1}. Work in this direction is in progress. We also point out that
noncommutativity is strictly related to the presence of a topological defect
on the edge of the bilayer system, which supports protected Majorana fermion
zero modes. That happens in close analogy with point defects in topological
insulators and superconductors, where the existence of Majorana bound states
is related to a $Z_{2}$ topological invariant \cite{teo1}.

We show the existence of the magic points of Ref.\cite{maldacena} drawing
parallels with the fixed points obtained by the group $SL\left( 2,Z\right)
/\Gamma _{2}$ which is obtained as residual symmetry after the m-reduction.
In addition, the effects of dissipation introduced by the two types of
localized defects are described by means of the GMT which are a realization of
the Wilson loop operators for this model.

We also want to emphasize that the introduction of a magnetic translation
for the neutral sector coupled to charged one is a peculiarity of our
twisted model.

The detailed calculation of the GMT action on the conformal blocks of the TM
model will be presented in a forthcoming publication as well as their
identification with Wilson loop operators of a pure Yang-Mills theory on a
two-torus \cite{AV3}.

Finally, we would stress that our work helps to shed new light on the
relationship between noncommutativity and QHF physics on one hand and
between string and $D$-brane theory and QHF physics on the other hand \cite%
{branehall2}. Recently we employed the $m$-reduction procedure in order to
describe non trivial phenomenology in different condensed matter systems
such as Josephson junction ladders and arrays \cite{noi3}, two-dimensional
fully frustrated $XY$ models \cite{noi} and antiferromagnetic spin-$1/2$
ladders with a variety of interactions \cite{noi7}. The role of
noncommutativity in these systems is now under study and will be the subject
of future publications.

\section{Appendix: TM on the torus}

The TM primary fields are composite operators and, on the torus, they are
described in terms of the conformal blocks (or characters) of the MR and the
Ising model \cite{cgm4}. The MR characters $\chi _{(\lambda ,s)}^{MR}$ with $%
\lambda =0,...2$ and $s=0,...,p$,\ are explicitly given by:
\begin{eqnarray}
\chi _{(0,s)}^{MR}(w|\tau ) &=&\chi _{0}(\tau )K_{2s}\left( w|\tau \right)
+\chi _{\frac{1}{2}}(\tau )K_{2(p+s)+2}\left( w|\tau \right)  \label{mr1} \\
\chi _{(1,s)}^{MR}(w|\tau ) &=&\chi _{\frac{1}{16}}(\tau )\left(
K_{2s+1}\left( w|\tau \right) +K_{2(p+s)+3}\left( w|\tau \right) \right)
\label{mr2} \\
\chi _{(2,s)}^{MR}(w|\tau ) &=&\chi _{\frac{1}{2}}(\tau )K_{2s}\left( w|\tau
\right) +\chi _{0}(\tau )K_{2(p+s)+2}\left( w|\tau \right) .  \label{mr3}
\end{eqnarray}
They represent the field content of the $Z_{2}$ invariant $c=3/2$ \ CFT \cite%
{MR} with a charged component ($K_{\alpha }(w|\tau )=\frac{1}{\eta (\tau )}%
\Theta \left[
\begin{array}{c}
\frac{\alpha }{4\left( p+1\right) } \\
0%
\end{array}
\right] \left( 2\left( p+1\right) w|4\left( p+1\right) \tau \right) $) and a
neutral component ($\chi _{\beta }$, the conformal blocks of the Ising
Model).

The characters of the twisted sector are given by:
\begin{eqnarray}
\chi _{(0,s)}^{+}(w|\tau ) &=&\bar{\chi}_{\frac{1}{16}}\left( \chi
_{(0,s)}^{MR}(w|\tau )+\chi _{(2,s)}^{MR}(w|\tau )\right)  \label{tw1} \\
\chi _{(1,s)}^{+}(w|\tau ) &=&\left( \bar{\chi}_{0}+\bar{\chi}_{\frac{1}{2}%
}\right) \chi _{(1,s)}^{MR}(w|\tau )  \label{tw2}
\end{eqnarray}
which do not depend on the parity of $p$;
\begin{eqnarray}
\chi _{(0,s)}^{-}(w|\tau ) &=&\bar{\chi}_{\frac{1}{16}}\left( \chi
_{(0,s)}^{MR}(w|\tau )-\chi _{(2,s)}^{MR}(w|\tau )\right)  \label{tw3} \\
\chi _{(1,s)}^{-}(w|\tau ) &=&\left( \bar{\chi}_{0}-\bar{\chi}_{\frac{1}{2}%
}\right) \chi _{(1,s)}^{MR}(w|\tau )  \label{tw4}
\end{eqnarray}
for $p$ even, and
\begin{eqnarray}
\chi _{(0,s)}^{-}(w|\tau ) &=&\bar{\chi}_{\frac{1}{16}}\left( \chi _{0}-\chi
_{\frac{1}{2}}\right) \left( K_{2s}\left( w|\tau \right) +K_{2(p+s)+2}\left(
w|\tau \right) \right)  \label{tw5} \\
\chi _{(1,s)}^{-}(w|\tau ) &=&\chi _{\frac{1}{16}}\left( \bar{\chi}_{0}-\bar{%
\chi}_{\frac{1}{2}}\right) \left( K_{2s+1}\left( w|\tau \right)
-K_{2(p+s)+3}\left( w|\tau \right) \right)  \label{tw6}
\end{eqnarray}
for $p$ odd.

Notice that only the symmetric combinations $\chi _{(i,s)}^{+}$ can be
factorized in terms of the $c=\frac{3}{2}$ \ and $c=\frac{1}{2}$ theory.
That is a consequence of the parity selection rule ($m$-ality), which gives
a gluing condition for the charged and neutral excitations.

Furthermore the characters of the untwisted sector are given by:
\begin{eqnarray}
\tilde{\chi}_{(0,s)}^{+}(w|\tau ) &=&\bar{\chi}_{0}\chi _{(0,s)}^{MR}(w|\tau
)+\bar{\chi}_{\frac{1}{2}}\chi _{(2,s)}^{MR}(w|\tau )  \label{vacuum1} \\
\tilde{\chi}_{(1,s)}^{+}(w|\tau ) &=&\bar{\chi}_{0}\chi _{(2,s)}^{MR}(w|\tau
)+\bar{\chi}_{\frac{1}{2}}\chi _{(0,s)}^{MR}(w|\tau )  \label{ut1} \\
\tilde{\chi}_{(0,s)}^{-}(w|\tau ) &=&\bar{\chi}_{0}\chi _{(0,s)}^{MR}(w|\tau
)-\bar{\chi}_{\frac{1}{2}}\chi _{(2,s)}^{MR}(w|\tau )  \label{vacuum2} \\
\tilde{\chi}_{(1,s)}^{-}(w|\tau ) &=&\bar{\chi}_{0}\chi _{(2,s)}^{MR}(w|\tau
)-\bar{\chi}_{\frac{1}{2}}\chi _{(0,s)}^{MR}(w|\tau )  \label{ut2} \\
\tilde{\chi}_{(s)}(w|\tau ) &=&\bar{\chi}_{\frac{1}{16}}\chi
_{(1,s)}^{MR}(w|\tau ).  \label{ut3}
\end{eqnarray}

\end{document}